\documentclass[useAMS,usenatbib]{mn2e}

\usepackage{bbm}
\usepackage{mathrsfs}

\renewcommand\>{\rangle}
\newcommand\beq{\begin{equation}}
\newcommand\eeq{\end{equation}}
\newcommand\bea{\begin{eqnarray}}
\newcommand\eea{\end{eqnarray}}
\newcommand\fr{\frac}
\newcommand\app{\approx}
\newcommand\X{\times}

\newcommand\cd{\cdot}
\newcommand\de{\delta}
\newcommand\ze{\zeta}
\newcommand\ph{\phi}
\newcommand\rms[1]{_{\mathrm{#1}}}
\newcommand\bk{\mathbf{k}}
\newcommand\br{\mathbf{r}}

\newcommand\cP{\mathcal{P}}
\newcommand\ai{a\rms{i}}
\newcommand\ar{a\rms{rh}}
\newcommand\ti{t\rms{i}}
\newcommand\trh{t\rms{rh}}
\newcommand\rhr{\rho\rms{r}}
\newcommand\rhv{\rho\rms{v}}
\newcommand\Ti{T\rms{i}}
\newcommand\Tph{T\rms{phys}}
\newcommand\Trh{T\rms{rh}}

\title[Large pre-inflationary
 thermal density perturbations]{Large pre-inflationary
 thermal density perturbations}
\author[Richard Lieu and T.W.B. Kibble]{Richard Lieu$^{1}$\thanks{E-mail:
lieur@uah.edu} and T.W.B. Kibble$^{2}$\\
$^{1}$Department of Physics, University of Alabama,
Huntsville, AL 35899, U.S.A.\\
$^{2}$Theoretical Physics Group, Blackett Laboratory, Imperial College, London SW7 2BZ}

\begin{document}

\maketitle

\label{firstpage}

\begin{abstract}
In some versions of the theory of inflation, it is assumed that before inflation began the universe was in a Friedmann-Robertson-Walker (FRW) stage, with the energy density dominated by massless particles.  The origin of the nearly scale-invariant density perturbations is quantum fluctuations in the inflaton field.  Here we point out that under those conditions there would necessarily also be large thermally induced density perturbations.  It is asserted that inflation would smooth out any pre-existing perturbations.  But that argument relies on linear perturbation theory of the scalar modes, which would be rendered invalid because of the non-negligibility of the vector and tensor modes when the perturbation in the total density becomes large.  Under those circumstances the original proof that inflation would have the desired smoothing effect no longer applies, {\it i.e.} for the theory to be robust an alternative (and hitherto unavailable) demonstration of the smoothing that takes account of these non-linear terms is necessary.

\end{abstract}


\section{Introduction}

The theory of inflation (\citet{gut81}, \citet{alb82}, \citet{lin82}) provides attractive solutions to a number of cosmological problems, including the large-scale homogeneity and flatness of the universe.  It has garnered strong support from the \texttt{COBE} and \texttt{WMAP} observations of the cosmic microwave background radiation (\citet{smo92}, \citet{spe07}, \citet{hin09}, \citet{kom11}).  Quantum fluctuations in a scalar inflaton field can explain the origin and near scale-invariant power spectrum of the primordial density perturbations, although getting the amplitude right requires fine tuning, as is establishing the preconditions for inflation.

Early versions of the theory envisaged a pre-inflationary Friedmann-Robertson-Walker (FRW) stage.  One of the claimed advantages of the theory was that the rapid expansion effectively erased all traces of that earlier phase, although it was realized before (\citet{fri84}) that that is not strictly true; what it does is not to eliminate perturbations but to stretch them to unobservable scales.  This leaves open the question of whether there could be perturbations on very tiny scales that are stretched to observable size.  It has been shown (\citet{mag07}) that without drastic modifications such perturbations could not explain the observed power spectrum of density perturbations.

Here we wish to go much further than \citet{mag07}, to argue that under rather general conditions the {\it same} initial small scale perturbations would indeed be {\it inevitable} and besides having the wrong power spectrum would be far too large in amplitude to be consistent with the use of linear perturbation theory.  This is because thermal radiation has large fluctuations on scales comparable with the thermal wavelength, and such scales have been stretched to cosmological size today.  Under these circumstances, it is not possible to say what the outcome would be.  It should be emphasized that there is a key difference between this paper and previous ones on the topics of `warm inflation' and `thermal inflation'.   These latter rely on either a mild to strong coupling between the inflaton and some other field to maintain a thermal medium even during the slow roll phase (\citet{arj96a,arj96b}), or an extension of standard model to include an {\it additional} period of inflation with a significant thermal component (\citet{bar96}).  The question we are asking, on the other hand, is `what happens if there was a pre-inflationary radiation dominated universe'?  This phase can certainly exist {\it independently} of `warm' or `thermal' inflation.  It is an especially pertinent question because the cosmologically relevant scales today could very easily have originated from scales well beneath one wavelength of such a radiation component.  None of the three papers cited above, nor papers related to them, addressed the extremely quantum phenomenon of fluctuations on the wavelength and sub-wavelength scales.  That is the purpose of our work.

It has also been previously argued (\citet{vac98}) that inflation requires fine-tuned initial conditions in the sense that the relevant part of the universe must already be very smooth.  We suggest that it is essentially impossible to satisfy that condition if there is a pre-inflationary FRW stage.

There are obvious ways in which the theory could avoid the problems we currently raise.  Most simply, the universe could be born inflating (\citet{vil82}, \citet{har83}).  Later versions of inflation, such as chaotic inflation (\citet{lin83}, \citet{lin86}) do not necessarily begin with a pre-inflationary FRW phase, and might therefore seem to be unaffected.  Even then, however, thermal radiation may be present, both because de Sitter spacetime has a horizon and because radiation is created by the decay of the slow-rolling inflaton.  This is an issue we hope to discuss in a future paper.

Let us assume that in the pre-inflationary phase the patch of the universe from which today's Hubble volume evolved contained a fluid of massless particles, with interactions weak enough to treat them as free, but strong enough to maintain thermal equilibrium.  The latter is confirmed, especially for the light particle field, by a detailed analysis of the thermalization rate during inflation, see \cite{bas13}.  We also assume the existence of an inflaton field providing vacuum energy that will eventually come to dominate and induce a period of inflation.

Now suppose that the vacuum energy starts to dominate at an initial time $\ti$, at which $\rhr(\ti)=\rhv(\ti)$, where $\rhr$ stands for the energy density of the radiation and $\rhv$ for that of the inflaton field.  When inflation ends at a reheating time $\trh$, the vacuum energy is converted during a relatively short time interval $\de t$ to radiation.  Thus $\rhr(\trh+\de t) \app \rhv(\trh)$.  Now $\rhv$ is nearly constant during the inflationary era, so $\rhv(\trh) \app \rhv(\ti)$.  Thus it follows that $\rhr(\trh+\de t) \app \rhr(\ti)$, which means that the physical temperature $\Tph$ of the initial radiation at $\ti$ is approximately the same as that of the newly created radiation just after $\trh$.  But note that this is not the \emph{same} radiation; the pre-existing radiation has by this time become a negligible fraction of the total energy.  This $\Tph$ is a large temperature, but still some orders of magnitude below the Planck energy.  It is convenient to use a `comoving temperature' $T=a\Tph$ which is nearly constant except during any period of warm inflation or reheating.  More precisely, $g^{1/3}T$ is constant, where $g$ is the number of helicity states of massless particles (with fermions counted as $\fr{7}{8}$).  For the universe since reheating, allowing for the effect of neutrino decoupling and electron-positron annihilation,
 \beq g^{1/3}T = \left(\fr{43}{11}\right)^{1/3}T_0, \eeq
where $T_0=2.7$K is the CMB temperature today.  Thus the approximate equality of the physical temperatures before and after inflation means that
 \beq \fr{\Ti}{\Trh} =\fr{T(\ti)}{T(\trh)}\app \fr{\ai}{\ar}
 = \fr{a(\ti)}{a(\trh)}=e^{-N}, \eeq
where $N$ is the number of $e$-folds of inflation (assuming there is no change in the value of $g$).  Equivalently, the comoving temperature in the pre-inflationary phase is
 \beq \Ti \sim 0.3 e^{-N}T_0. \label{Ti}\eeq
This assumes that $g \sim 100$, but the result is not very sensitive to the value of $g$.

\section{Density fluctuations in thermal equilibrium}

In any thermal system there are fluctuations.  Before proceeding further, we need to estimate how big the density fluctuations are in a gas of massless particles in thermal equilibrium.  Of course in the classical region where $k \ll T$, the answer is well known and easily derivable from thermodynamic arguments (see e.g.~\citet{fer08}).  But the result for $k\geq T$ is not so obvious.

Since we are assuming that the system is expanding slowly enough to remain in thermal equilibrium, it will be sufficient to consider Minkowski spacetime.  For simplicity we consider only a single massless scalar field $\ph$.  For the general case, we simply have to multiply by the number of helicity states $g$.

The energy density of the field is $u(t,\br)=\frac{1}{2}:[\dot\ph^2(t,\br)+(\nabla\ph)^2(t,\br)]:$, where the colons denote normal ordering.  In terms of creation and annihilation operators,
 \bea u(0,\br)&=&\frac{1}{2} \int \fr{d^3\bk}{(2\pi)^3 2k}
 \int \fr{d^3\bk'}{(2\pi)^3 2k'}
 (kk'+\bk\cd\bk')\nonumber\\
 &&\ \big[-a(\bk)a(\bk')e^{i(\bk+\bk')\cd\br} -\;a^*(\bk)a^*(\bk')e^{-i(\bk+\bk')\cd\br}\nonumber\\
 &&\ +2a^*(\bk)a(\bk')e^{i(\bk-\bk')\cd\br}\big]. \label{u}  \eea
Now in equilibrium at temperature $T$, $\<a^*(\bk)a(\bk')\>=16\pi^3 kn(k)\de_3(\bk-\bk')$, where $n(k)=(e^{k/T}-1)^{-1}$.  For the mean value of $u$ only the last term in (\ref{u}) contributes, so
 \beq \<u\> = \fr{1}{(2\pi)^3} \int d^3\bk\, kn(k) =
 \fr{1}{2\pi^2}\int_0^\infty k^3 n(k)dk
 = \fr{\pi^2}{30}T^4. \label{uav}  \eeq
Of course the result for photons is the same, except for a factor 2 arising from the two polarization states.

What we want to calculate is the dimensionless density perturbation power spectrum,
 \beq \cP(k)=\fr{k^3}{2\pi^2}P(k)
 = \fr{k^3}{2\pi^2}\int d^3\br\,\xi(\br)e^{-i\bk\cd\br},
 \label{De}  \eeq
where $\xi$, the density correlation function, is defined by
 \beq \xi(\br) = \fr{\<u(0,\br) u(0,{\bf 0})\>}{\<u\>^2}-1,
 \label{xi}  \eeq
To evaluate this, we now substitute (\ref{uav}) into $\<u(0,\br) u(0,{\bf 0})\>$ twice.  Clearly, the only terms that will give nonzero contributions are those with two $a$s and two $a^*$s.  Moreover, since there is no correlation between different wave vectors, there must be two delta functions relating the arguments of the $a$s to those of the $a^*$s.  One term will simply reproduce the square of $\<u\>$, so cancels the 1 in (\ref{xi}).  In addition, there is a divergent term that would be present even at zero temperature, which must also be subtracted.  In the remaining terms, the integration variables can be chosen so that each contains the same delta function $\de_3(\bk-\bk_1-\bk_2)$:
 \bea \cP(k)&\!\!\!=\!\!\!&
 \fr{k^3}{2(2\pi)^5\<u\>^2}\int \fr{d^3\bk_1}{k_1}
 \int \fr{d^3\bk_2}{k_2} \de_3(\bk-\bk_1-\bk_2)\nonumber\\
 &&\  n(k_1)[ n(k_2)+1] \X\;[(k_1k_2+\bk_1\cd\bk_2)^2+\nonumber\\
 &&\ (k_1k_2-\bk_1\cd\bk_2)^2].
 \label{Pk}
 \eea
Next we can do the angular integrations, using $\bk_1\cd\bk_2=\frac{1}{2} (k^2-k_1^2-k_2^2)$.  It is convenient to introduce the two dimensionless integration variables $X=(k_1+k_2)/k$ and $Y=(k_1-k_2)/k$.  Thus we find
 \bea \cP(k) &=& \fr{k^8}{256\pi^4\<u\>^2}
 \int_1^\infty dX \int_{-1}^1 dY \, [(X^2-1)^2+(Y^2-1)^2] \X \nonumber\\
 &&\ [e^{(X+Y)k/2T}-1]^{-1} [1-e^{-(X-Y)k/2T}]^{-1} .
 \eea
To perform the integrals, we may expand the denominators, obtaining
 \bea \cP(k) &\!\!\!=\!\!\!& 900\left(\fr{k}{2\pi T}\right)^8
 \sum_{r=1}^\infty \sum_{s=0}^\infty
 \int_1^\infty dX\, e^{-(r+s)Xk/2T} \int_{-1}^1 dY\, \X \nonumber\\
 &&\  e^{-(r-s)Yk/2T} \X\;[(X^2-1)^2+(Y^2-1)^2].
 \eea
(Note the differing ranges of the summations.  The $s=0$ term arises from the $+1$ in (\ref{Pk}).)  The integrals can now be performed.  To write the result in a reasonably concise form, we introduce the abbreviation
 \beq f_r = \fr{rk}{2T}, \eeq
Then we find
 \bea \cP(k) &\!\!\!=\!\!\!& \fr{225}{8}
 \left(\fr{k}{\pi T}\right)^8
 \sum_{r=1}^\infty \sum_{s=0}^\infty
 \bigg\{\fr{e^{-f_{r+s}}(f_{r+s}^2+3f_{r+s}+3)}{f_{r+s}^5}\; \X \nonumber\\
 &&\ \fr{e^{f_{r-s}}-e^{-f_{r-s}}}{f_{r-s}}
 +\;\fr{e^{-f_{r+s}}}{f_{r+s} f_{r-s}^5}\;
 [e^{f_{r-s}}(f_{r-s}^2-3f_{r-s} \nonumber\\
 &&\ +3) - e^{-f_{r-s}}(f_{r-s}^2+3f_{r-s}+3)] \bigg\}.\label{P}
 \eea

We are most interested in the limiting forms of $\cP$ for small and large $k/T$.  Expanding in powers of $k/T$ we find that all terms in $(k/T)^2$ cancel, and the leading term is
 \beq \cP(k)\sim \fr{5400 k^3}{\pi^8 T^3}
 \sum_{r=1}^\infty \sum_{s=0}^\infty
 \fr{1}{(r+s)^5}
 = \fr{5400 k^3}{\pi^8 T^3}\ze(4)=\fr{60k^3}{\pi^4 T^3}, \label{PSmallk}  \eeq
valid for $k \to 0$ and in agreement with the classical expression for this limit.  To further elaborate upon this agreement, (\ref{PSmallk}) is smaller than the power spectrum of thermal photons in the small $k$ limit by a factor of two, which is exactly what one expects because photons have two internal degrees of freedom ({\it i.e.}~$(\de u)^2$ and $\bar u$ are both larger, hence $(\de u/\<u\>)^2$ smaller, by two times relative to the massless particles).

So far as the large-$k$ behavior is concerned, we can drop all terms involving decreasing exponential factors.  The only surviving terms are those with $s=0$ and where the two exponentials cancel.  Keeping only these, we have
 \bea \cP(k) &\!\!\!\sim\!\!\!& \fr{225}{8}
 \left(\fr{k}{\pi T}\right)^8
 \sum_{r=1}^\infty
 \fr{f_r^2+3}{f_r^6} \qquad(k\to\infty). \label{PLargek}
 \eea
The leading term is the one in $1/f_r^4$, which gives
 \beq \cP(k)\sim \fr{900\ze(4)k^4}{\pi^8 T^4}=\fr{10 k^4}{\pi^4 T^4}
 \qquad (k\to\infty). \label{DeLargek}  \eeq
The point at which the two asymptotic forms (\ref{PSmallk}) and (\ref{PLargek}) agree is where $k/T=6$.  It would also appear that, from the nature of this calculation, $P(k)$ becomes large at large $k$ for many other states besides exactly thermal ones.

Before leaving this section we revisit our procedure of divergence removal to show that it is robust and consistent.  When calculating $\<u\>$ we subtracted the
infinite vacuum contribution, or equivalently normally ordered the factors
in $u$.  In the case of $\<u(\br) u({\bf 0})\>$ we again subtracted the
divergent vacuum contribution, but in that case the procedure is {\it not}
equivalent to normal ordering.  In fact, although the direct normal ordering of $\<u(\br) u({\bf 0})\>$ does lead to a $e^{-k/T}$ cutoff in $\cP(k)$ in the $k \gg T$ limit, in the $k \ll T$ limit the $\cP(k)$ so produced disagrees with the well-known formula for the thermal density fluctuations of a massless scalar.  Specifically, the effect of $:\<u(\br) u({\bf 0})\>:$ is to replace $n(k_2)+1$ in eq.~(8) by $n(k_2)$.  In eq.~(10) that would mean dropping the $s=0$ term in the summation.  The result in eq.~(13) would be to replace $\ze(4)$ by $\ze(4)-\ze(5)$, giving the wrong coefficient.  Concerning the well known behavior of thermal fluctuations on super-wavelength scales, therefore, directly normally ordering $\<u(\br) u({\bf 0})\>$ would lead to the wrong answer altogether.

To elaborate further the physical meaning of our renormalization procedure, we note that as with any bosonic field theory, the vacuum expectation value of the
energy density is formally divergent.  Unless one invokes a manifestly supersymmetric formalism, wherein the
cancelation between fermionic and bosonic contributions happens automatically, one must explicitly get rid of such infinities (which is the common practice).  Here we achieve that by
subtracting the divergent zero-temperature expectation value.  This is
equivalent to normal ordering the factors.  Similarly when it comes to
calculating the correlation function, there is again a divergent
zero-temperature contribution that must be removed, although in that
case the procedure is {\it not} equivalent to normal ordering, rather the same
as the one was adopted by {\it e.g.} \cite{hal04}, eq. (44), although this earlier paper did not include any derivation of the power spectrum in the $k \gg T$ reg\'ime.

\section{Perturbations in the expanding universe}

Thus far we have considered a static universe.  Suppose, however, that we have an expanding universe in which expansion is slow enough to allow the radiation to remain close to thermal equilibrium.  The results above will then hold to a good approximation.  Note that since both $k$ and $T$ are `comoving', $k/T=k\rms{phys}/T\rms{phys}$.

The result (\ref{DeLargek}) suggests that $\cP(k)$ becomes much larger than unity when $k\gg T$.  This means that the standard deviation of a measurement of the radiation energy in a volume of size smaller than the typical wavelength of the radiation is much larger than its mean value.  Given that the energy cannot be negative, this implies a highly non-gaussian distribution.  Clearly it would be difficult to devise an experiment to measure directly the energy density perturbations on scales smaller than this wavelength.  We are interested in a `measurement' mediated by perturbations in the gravitational field.

Moreover, we are discussing a \emph{quantum} fluctuation in the energy of the radiation gas.  There is an inevitable ambiguity because we are interested in \emph{classical} gravitational effects of a \emph{quantum} source.  One would normally assume that there is a process of decoherence, induced by the exponential expansion, in which the quantum distribution is replaced by a classical statistical distribution.  In estimating the magnitude of the density perturbations as the universe expands, we might use a `typical' value of $\de$.  Mention should also be made of the fact that, for $k > T$, $\sqrt{\cP(k)}$ cannot actually be regarded as a `typical' value of $\de$ on the length scale $1/k$.

\section{Discussion}

We are interested in the dimensionless power spectrum $\cP(k)$ of the density fluctuations.  Here $k$ and $T$ are both comoving, so $k/T$ is the same as in physical units.  So long as $k\ll T$, then $\cP(k)$ is given by (\ref{PSmallk}).  But the key point is that when $k/T$ becomes of order unity, or in other words when the wavelength of interest becomes of the same order as the typical wavelength of the radiation then $\cP(k)$ itself is of order 1.  According to the above calculation, for $k\gg T$ then we have even larger perturbations.  Consequently we cannot rely on linear theory (while this point needs no further elaboration by now, it is nevertheless further discussed in \citet{lie13}).

The problem here is that the relevant scale can easily be one of cosmological significance.  According to (\ref{Ti}), $k\sim T$ when
$1/k \sim 3e^N/T_0$.
Thus for $N=60$ this scale would be a few Mpc.

It has been argued that any pre-existing perturbations present before the start of inflation will be ironed out by the rapid expansion (e.g.~\citet{bar83}).  However, the argument fails if these perturbations are of order unity, because then linear perturbation theory is invalid.  Proceeding with the conventional approach ({\it i.e.}~ignoring the effects discussed in the end of the previous section), it might be thought sufficient for the validity of perturbation theory that the gravitational potential $\Phi\app Ga^2\de\rho/k^2 \ll 1$ (this same inequality holds also for the radiation era superseding reheating, which explains why sub-wavelength scale fluctuations of ordinary radiation does not produce small black holes).  However, when the relative {\it density} excursion is of order 1 or larger, as is the case on the scale $k\sim T$ at the start of inflation, then even with $\Phi \ll 1$ the vector and tensor modes are coupled to the scalar modes and become important (see section 4.6 of \cite{ber93}, also \cite{mol97}).  Although it is possible that nonlinear effects can smooth out the perturbations, this has not been demonstrated.  It is far more likely that they would be observationally unacceptable.

One `escape' might be to postulate a substantially larger value of $N$, say $N \geq 70$.  In that case the length scale at which $k\sim T$ would be larger than the present Hubble radius.  In reality this `solution' actually accentuates the problem,  because according to (\ref{DeLargek}) $\cP(k)$ has not decreased towards the smaller cosmological scales that corresponded to $k > T$ (this understates the case, as $\cP(k)$ actually increases), {\it i.e.} on these scales linear perturbation is invalidated even more.   In fact, under the scenario patches of size such that $k > T$ will inflate at very different times, due to the large variation in the density ratio of radiation to inflaton, resulting in a highly inhomogeneous observable universe.  Another apparent way out is to invoke a protracted reheating period at the end of inflation, so that the physical temperature before inflation is higher than after, thereby placing the thermal wavelength scale of $k \sim T$ deeper within the horizon at $t=t_i$.  Yet there is a compensating effect at work against this: if reheating takes many Hubble times to complete, the horizon will move out significantly during this phase, {\it i.e.} the initial horizon at $t=t_i$ will likewise be smaller, rendering it difficult to `bury' the problematic $k \geq T$ scales.


The issue raised poses a real difficulty for the theory of inflation.  The scenario we have discussed is only one possibility.  It could also be that the pre-inflationary phase is dominated by extremely massive cold particles, such as monopoles arising from an early phase transition, but in such a case it seems likely that there will again be large fluctuations on very small scales, unless a superfluid is postulated for the phase (as noted after (\ref{DeLargek}), most fluids would exhibit a $\cP (k)$ divergence towards $k \gg T$, not just thermal ones).  Even if inflation is `warm' it is not obvious how the initial non-perturbative situation of the pre-inflationary phase may be averted.  However, if such a phase is altogether absent, or indeed consists of a highly homogeneous superfluid to start with, then of course that would be the `ultimate fix'.  But in this case, inflation cannot claim to have {\it explained} why we fail to find evidence (directly or indirectly) for the relic particles and the curvature of space.

\section*{Acknowledgments}

We wish to acknowledge helpful correspondence with Massimo Giovannini, Michael Joyce, Joao Magueijo, Ian Moss and David Wands, as well as critical comments from Andrei Linde and David Lyth.

\end{document}